\documentclass[a4paper,10pt,preprint,twocolumn,3p]{elsarticle} 
\usepackage{lineno}
\usepackage{hyperref}
\usepackage{amsmath,amssymb}
\usepackage{graphicx}
\usepackage{braket}
\usepackage{float}
\usepackage{comment}
\biboptions{sort&compress}
\modulolinenumbers[5]

\journal{Physics Letters B}

\begin{document}

\begin{frontmatter}

\title{Quasiparticle pairing encoding of atomic nuclei for quantum annealing}

\author[inst1,inst2]{Emanuele Costa\corref{cor1}}
\ead{emanuele.costa@icc.ub.edu}

\author[inst3]{Axel Pérez-Obiol}
\ead{axel.perezobiol@uab.cat}

\author[inst1,inst2]{Javier Menéndez}
\ead{menendez@fqa.ub.edu}

\author[inst1,inst2]{Arnau Rios}
\ead{arnau.rios@fqa.ub.edu}

\author[inst4,inst5]{Artur García-Sáez}
\ead{artur.garcia@bsc.es}

\author[inst1,inst2]{Bruno Juliá-Díaz}
\ead{bruno@fqa.ub.edu}

\address[inst1]{Departament de Física Quàntica i Astrofísica, Universitat de Barcelona, 
08028 Barcelona, Spain}
\address[inst2]{Institut de Ciències del Cosmos, Universitat de Barcelona, 
08028 Barcelona, Spain}
\address[inst3]{Departament de Física, Universitat Autònoma de Barcelona, 08193 Bellaterra, Spain}
\address[inst4]{Barcelona Supercomputing Center, 08034 Barcelona, Spain}
\address[inst5]{Qilimanjaro Quantum Tech, 08019 Barcelona, Spain}

\cortext[cor1]{Corresponding author.}

\begin{abstract}
Quantum computing is emerging as a promising tool in nuclear physics. However, the cost of encoding fermionic operators hampers the application of algorithms in current noisy quantum devices. In this work, we analyze an encoding scheme based on pairing nucleon modes. This approach significantly reduces the complexity of the encoding, while maintaining a high accuracy for the ground states of semimagic nuclei across the $sd$ and $pf$ shells and for tin isotopes. In addition, we also explore the encoding ability to describe open-shell nuclei within the above configuration spaces. When this scheme is applied to a trotterized quantum adiabatic evolution, our results demonstrate a computational advantage of up to three orders of magnitude in CNOT gate count compared to the standard Jordan-Wigner encoding. Our approach paves the way for efficient quantum simulations of nuclear structure using quantum annealing, with applications to both digital and hybrid quantum computing platforms.
\end{abstract}


\end{frontmatter}

\section{Introduction}
Quantum computing exploits the unique ability of quantum systems to process and spread information. Its wide range of applications covers optimization, machine learning, computational chemistry or fundamental high-energy physics~\cite{Zaman2023SurveyQML,McArdle2020QuantumComputationalChemistry,Sajjan2022QuantumML,Bauer:2023qgm}, but quantum computing is particularly promising in simulating quantum systems~\cite{Daley2022}, a domain where classical approaches face severe computational barriers.
Several classical algorithms --such as quantum Monte Carlo, neural-network quantum states, and tensor network methods-- have been designed to overcome these limitations~\cite{annurev:/content/journals/10.1146/annurev-conmatphys-033117-054307,lange2024architectures,Montangero2018,Orus2019}, yielding promising but not universally satisfactory results. 

In particular, in low-energy nuclear physics the interactions between the relevant degrees of freedom, protons and neutrons, are particularly complex. Therefore, precise classical simulations remain highly challenging. In recent years, quantum computing has emerged as a promising novel avenue to address these challenges. In nuclear structure, for instance, several quantum algorithms for ground- and excited-state simulation have been proposed, including those based on variational quantum eigensolvers (VQEs) \cite{Dumitrescu2018,Siwach:2021tym,PhysRevC.105.064317,PhysRevC.106.034325,Stetcu:2021cbj,perez2023nuclear,Sarma:2023aim,Bhoy_2024,Perez-Obiol:2024vjo,zhang2025excited,Gu:2025yww}, projective approaches \cite{PhysRevC.105.024324,PhysRevC.108.L031306}, imaginary time evolution~\cite{Li:2023qdd} or quantum adiabatic evolution methods \cite{costa2024quantum}. These works yield promising results in small-scale classical simulations. In addition, other proposed methods focus on nuclear dynamics~\cite{Weiss:2024mie,Singh:2025ubp,ZHANG2025139187} or scattering theory~\cite{PhysRevC.109.064623,PhysRevC.111.034001}. 

Despite the natural advantage of quantum algorithms in simulating nuclear systems, their practical implementation in digital quantum computers usually demands deep quantum circuits with a large number of non-local and non-Clifford gates. These requirements compromise their feasibility on current quantum devices due to their relatively small size and significant noise~\cite{Bharti2021_bitter_truth}. While error mitigation techniques can suppress the effects of noise, they remain limited, particularly in deep circuits or at high noise levels \cite{Czarnik2021errormitigation,kandala2019error,PhysRevLett.119.180509,PhysRevX.8.031027}. Likewise, analog quantum simulators are constrained by hardware design and can efficiently simulate only very specific classes of quantum systems \cite{PRXQuantum.2.017003,RevModPhys.86.153,Cirac2012QuantumSimulation} that do not include the complex, non-local, nuclear Hamiltonian. 

In nuclear-structure applications, most of the computational overhead, quantified by circuit depth, arises from encoding the fermionic operators associated with protons and neutrons into the qubit representation. In general, transformations such as Jordan-Wigner (JW)~\cite{jw} and Bravyi-Kitaev~\cite{bk} lead to long strings of non-local Pauli operators, which are difficult to implement efficiently on both digital and analog platforms \cite{Seeley2012,Cao2019}. 
Several alternative mappings \cite{levy2022towards,derby2021compact,whitfield2016local,streif2019solving,PhysRevA.91.012315,babbush2014adiabatic,Gray-code-1} have been proposed to address this issue, but most of them are tailored to specific lattice geometries or symmetries and cannot be easily generalized to nuclear physics. Nonetheless, two recent works focused on nuclear structure present promising strategies~\cite{Gray-code-2,Stevenson25}.

In condensed matter physics and quantum chemistry a mapping of fermions into hardcore bosons has demonstrated a reduction in the quantum computational cost of algorithms such as VQE~\cite{10.1063/1.3613706,10.1063/1.4904384,doi:10.1021/acs.jpclett.2c00730,del2025hybrid,Krompiec:2025fac}. In this letter, we follow this idea and adopt a full-hardcore boson mapping to reduce the high computational cost to obtain nuclear-shell-model (NSM) ground states using quantum algorithms~\cite{Stetcu:2021cbj,perez2023nuclear,Li:2023eyg,costa2024quantum,Carrasco2024}. We note that Yoshida {\it et al.} have used a similar mapping in a study restricted to two-valence-neutron systems~\cite{PhysRevC.109.064305}. 

In our work, we encode the Hamiltonian in terms of quasiparticles constructed from pairs of like-nucleon modes with opposite magnetic quantum numbers. 
These quasiparticles obey hardcore boson statistics \cite{doi:10.1142/S0217979206034947,10.21468/SciPostPhysLectNotes.82} and can thus be directly mapped onto qubit degrees of freedom. We then systematically analyze, in medium-mass nuclei, whether the effective Hamiltonian that results from this encoding accurately captures nuclear ground-state properties. 
Finally, we demonstrate that our approach reduces the computational overhead associated with the mapping by several orders of magnitude, as quantified by the quantum resource requirements for a trotterized adiabatic time evolution (T-QA) on a digital quantum computer~\cite{costa2024quantum}.


\section{Method}
\label{Method}

The NSM~\cite{RevModPhys.77.427,RevModPhys.92.015002,annurev:/content/journals/10.1146/annurev.ns.38.120188.000333,annurev:/content/journals/10.1146/annurev-nucl-101917-021120} describes nuclear structure based on protons and neutrons interacting through a valence-space Hamiltonian:
\begin{equation}
    H_\text{NSM}=\sum_a e_a c^{\dagger}_a c_a+ \frac{1}{4}\sum_{abcd} v_{abcd}c^{\dagger}_a c^{\dagger}_b c_d c_c,
    \label{eq:NSM}
\end{equation}
where $e_a$ are single-particle energies, and $v_{abcd}$ two-body matrix elements. These can be obtained solely from chiral Hamiltonians~\cite{annurev:/content/journals/10.1146/annurev-nucl-101917-021120,PhysRevC.91.064301,PhysRevLett.113.142502} or by including some phenomenological adjustments~\cite{RevModPhys.77.427,RevModPhys.92.015002}. Here, $c^{\dagger}_a$ and $c_a$ stand for fermion creation and annihilation operators, respectively, for the nucleon mode $a$. Each mode $a$ represents a single-particle state in the valence space, and it is described by the principal quantum number $n_a$, the total angular momentum $j_a$, its third component $m_a$, and the isospin $t_{a}=1/2$ with third component $t_{z,a}$. 

For the NSM, the many-body basis comprises the set of Slater determinants $\ket{\mathbf{s}}$, represented as bitstrings $\mathbf{s}=(0,1,0,..,1,0)$, where 1's indicate occupied states, and 0's empty ones. For an isotope with $N_n$ neutrons and $Z_p$ protons in the valence space, the dimension of the many-body basis is $\dim(\mathcal{H}) = \binom{D}{Z_p} \times \binom{D}{N_n}$. Here, the total number of nucleon modes (or the number of entries in the bitstring $\mathbf{s}$) is $2D = 2\sum_{j \in J} (2j+1)$, with $J$ the set of orbitals which group single-particle states that only differ in $m_a$, given in spectroscopic $n l_j$ notation. We restrict our simulations to basis states with total magnetic quantum number $M = 0$. In our study, we consider three configuration spaces: the \emph{sd} shell ($D=12$), containing proton and neutron $0d_{5/2}$, $1s_{1/2}$ and $0d_{3/2}$ orbitals; the \emph{pf} shell ($D=20$), covering the orbitals $0f_{7/2}$, $1p_{3/2}$, $0f_{5/2}$, and $1p_{1/2}$ for both kind of nucleons; and the valence space ($D=32$) comprising the neutron $0g_{7/2}$, $1d_{5/2}$, $1d_{3/2}$, $2s_{1/2}$, and $0h_{11/2}$ orbitals.

\subsection{Quasiparticle pairing encoding}

Nuclear pairing plays a key role among the various components of the nuclear force~ \cite{RevModPhys.75.607,Rios2017_PairingSRC} and is capital for nuclear structure. It describes the attractive force between nucleons with the same isospin projection and opposite projection of the total angular momentum $m$. Indeed, this mechanism underpins several nuclear theory frameworks, including the seniority model and the interacting boson model~\cite{talmi_book,suhonen2007nucleons,Iachello1987}. In this spirit, we define the following quasiparticle operator,
\begin{equation}
    Q^{\dagger}_A = c^{\dagger}_{a} c^{\dagger}_{\tilde{a}},
\end{equation}
where $\tilde{a}$ denotes the time-reversed partner of mode $a$ with $m_{\tilde{a}}=-m_a$, while sharing the same set of $\lbrace n_a, j_a, t_{z,a}\rbrace$ quantum numbers. The label $A$ indicates the quasiparticle mode $\lbrace j_a,t_{z,a},m_a,-m_a \rbrace$.
These quasiparticle operators obey commutation relations consistent with a hardcore boson algebra \cite{doi:10.1142/S0217979206034947,10.21468/SciPostPhysLectNotes.82}, {\it i.e.},
\begin{align}
    [ Q_A,Q^{\dagger}_B]&=\delta_{A,B}(1-n_{a}-n_{\tilde{a}}), \\
    [Q_A,Q_B]&=[Q^{\dagger}_A,Q^{\dagger}_B]=0.
\end{align}
This algebra reflects the underlying fermionic structure and the fact that each pair can be occupied only once. The quasiparticle operators can be directly mapped onto qubit operators using the correspondence:
\begin{align}
    Q^{\dagger}_A&=S^{+}_A \,,  \\ 
    Q_B&=S^{-}_B \,, \\ 
    N_A&=Q^{\dagger}_A Q_A=(1+Z_A)/2\,, 
\end{align}
where $S^{+}=\frac{1}{2}(X+iY)$, $S^{-}=\frac{1}{2}(X-iY)$ and $X,Y,Z$ are the Pauli matrices. With this mapping, the label $A$ now corresponds to the qubit index of the quantum circuit. This directly avoids the need for fermion-to-qubit transformations like in the JW or Bravyi-Kitaev mappings.

For the quasiparticle states, we adopt an ordering based on the single-particle energy of the corresponding nucleon modes. For example, focusing on neutrons in the $sd$ shell, we start from the lowest-energy pairs that can be formed with $0d_{5/2}$ states, and assign the highest labels to the pairs composed by neutrons in the $0d_{3/2}$ orbital. The number of quasiparticle modes per orbital $j_a$ is half the number of magnetic states, $j_a+1/2$. For example, the $0d_{5/2}$ orbital yields three modes, while the $1s_{1/2}$ and $0d_{3/2}$ orbitals contribute one and two, respectively. Therefore, this reformulation not only bypasses complex fermion mappings, but also halves the number of qubits.

%
\begin{figure*}[t]
    \centering
    \includegraphics[width=1\linewidth]{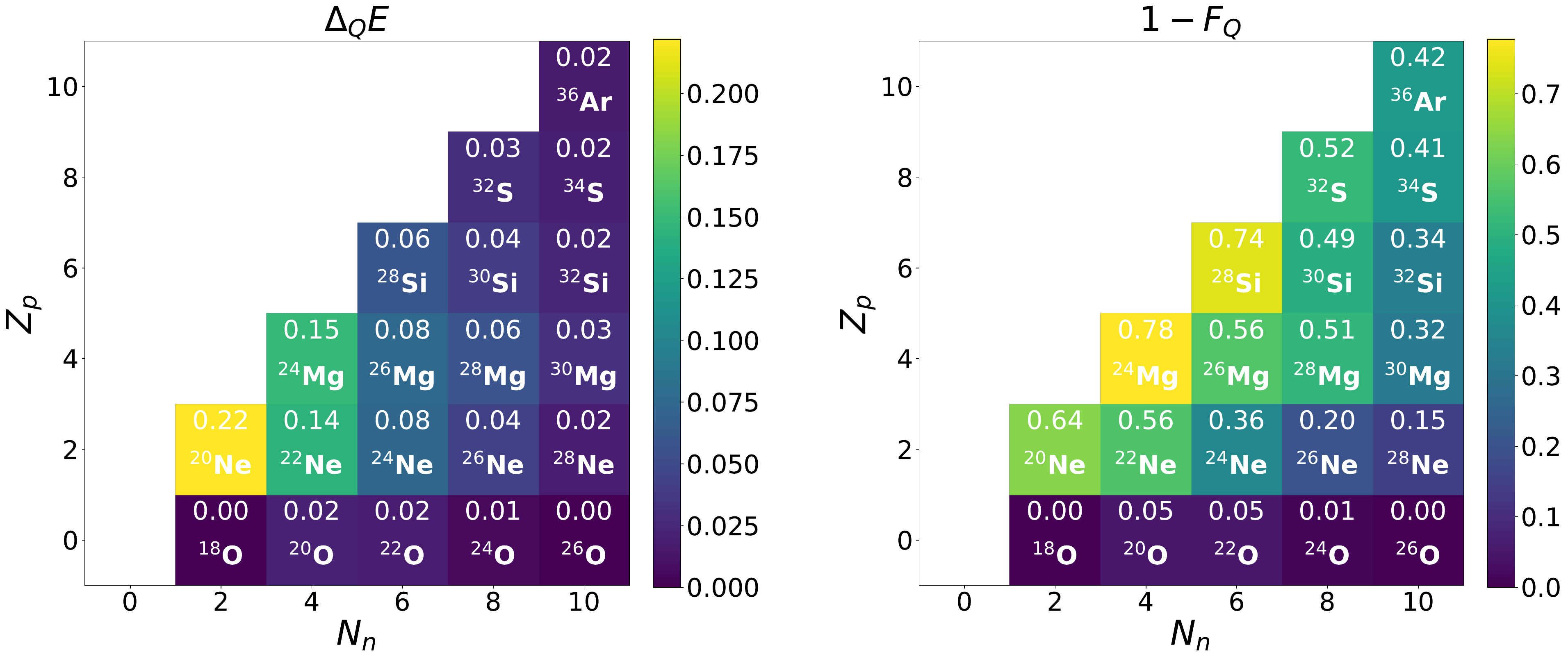}
    \caption{Energy relative error $\Delta_Q E$ (left panel) and infidelity $1-F_Q$ (right panel) of the ground state obtained with $H_Q$ with respect to the exact FCI result, for nuclei across the $sd$ shell in terms of their number of protons ($Z_p$) and neutrons ($N_n$) in the valence space. Darker colors indicate more accurate results obtained with $H_Q$.
    }
    \label{fig:1}
\end{figure*}

The nuclear Hamiltonian thus needs to be written in terms of the new degrees of freedom. We build the quasiparticle Hamiltonian by projecting the original NSM Hamiltonian of Eq.~\eqref{eq:NSM} into the many-body basis of quasiparticle pairs,
\begin{align}
H_Q=\mathbb{Q}H\mathbb{Q}\,.
\end{align}
Here, $\mathbb{Q}$ is the projector onto the space spanned by Slater determinants composed exclusively of quasiparticle modes, denoted as $\ket{\mathbf{S}_Q}=\prod_A Q^{\dagger}_A\ket{\Omega}$, where $\ket{\Omega}$ is the core of the NSM calculation. The complementary projector $\mathbb{R}=\mathbb{I}-\mathbb{Q}$ defines the subspace of many-body states that do not correspond to pure quasiparticle states. For nuclear ground states where the physics of pairing is dominant, we expect $H_Q$ to capture the relevant correlations of the system. In contrast, if the contributions from $\mathbb{R}$ dictate the nuclear structure --for example, in nuclei where deformation plays a key role-- important correlations are not captured in the projected Hamiltonian, and the description of the ground state in terms of quasiparticles loses accuracy. 

In this sense, $H_Q$ can be viewed as the zeroth-order term of a Brillouin-Wigner expansion in the projectors 
$\mathbb{Q}$ and $\mathbb{R}$ \cite{Hubac_2000,brillouin3,brillouin1,HubacWilson2009_BWMethods,brillouin2}. Higher-order terms would systematically reintroduce the virtual effects of $\mathbb{R}$-space configurations, thereby recovering physics beyond the pure quasiparticle picture.
Since the nuclear Hamiltonian contains at most two-body terms in the original fermionic operators, $H_Q$
contains up to four-qubit terms. Specifically, to lowest order, the resulting spin Hamiltonian reads
\begin{align}
    H_Q&=\frac{1}{2}\sum_{AB} g_{AB} S^{+}_A S^{-}_B \nonumber \\
    &+ \frac{1}{4}\sum_{ABCD} g_{AB}^{CD} S^{+}_A S^{+}_B S^{-}_C S^{-}_D,
    \label{eq h_q qubit}
\end{align}
where $g_{AB}$ are one-body quasiparticle matrix elements, and $g_{AB}^{CD}$ are irreducible two-body terms (all two-body terms that cannot be described by a combination of one-body terms). We compute these matrix elements numerically via matrix projection, but they may also be computed analytically. For example, the one-body coupling $g_{AB}$ takes the form
\begin{equation}
g_{AB}=\left( e_{a}+e_{\tilde{a}}\right)\left(\delta_{a,b}\delta_{\tilde{a},\tilde{b}}
-\delta_{\tilde{a},b} \delta_{a,\tilde{b}} \right)-v_{a,\tilde{a},b,\tilde{b}}.
\end{equation}
For the two-body quasiparticle interaction, due to the two-body nature of the nuclear Hamiltonian, only diagonal terms contribute, and $g_{AB}^{CD}$ simplifies, 
\begin{equation}
    g_{AB}^{AB}=v_{b,a,b,a}+v_{\tilde{a},b,\tilde{a},b}+v_{a,\tilde{b},a,\tilde{b}}+v_{\tilde{a},\tilde{b},\tilde{a},\tilde{b}}\,.
\end{equation}
%

\subsection{Implementation for quantum annealing}

The key advantage of the quasiparticle mapping lies in the simplification of the NSM Hamiltonian. This enables operators such as time evolution to be implemented with substantially fewer two-qubit gates. To explore this aspect, we benchmark the quasiparticle encoding by simulating a T-QA evolution and compare its performance with the standard JW mapping of $H_\text{NSM}$. 

The time-dependent interpolating Hamiltonian is given by
\begin{align}
    H(t) = \left(1 - \frac{t}{\tau}\right) H_D + \frac{t}{\tau} H_T,
\end{align}
where \( \tau \) is the total evolution time. 
For the standard fermion implementation of the NSM, the driver Hamiltonian $H_D$ is defined in Ref.~\cite{costa2024quantum}.
In the quasiparticle basis, instead, it takes the form
\begin{equation}
    H_D = \frac{E_0}{2(N_n + Z_p)} \sum_A S_Q[A] (1 - Z_A),
\end{equation}
where $ E_0 = \bra{\mathbf{S}_Q} H_Q \ket{\mathbf{S}_Q} $ is the energy of the initial quasiparticle Slater determinant $ \ket{\mathbf{S}_Q} $. This element of the many-quasiparticle basis is selected so that it is built with the nucleon modes with lowest single-particle energy, and among them, those with largest value of $m_a$. This is the same prescription as in Ref.~\cite{costa2024quantum}. The target Hamiltonian $H_T$ is the standard NSM Hamiltonian of Eq.~\eqref{eq:NSM} in the fermonic simulations, or the pairing encoded version of Eq.~\eqref{eq h_q qubit} in the quasiparticle simulations. In both cases, we implement the time evolution using a second-order Suzuki-Trotter decomposition of the unitary operator,
$
U(\Delta t + t, t)\,,
$
with a timestep chosen such that $ \Delta t \, \omega = 0.1 $, where $ \omega $ is the energy scale $1$ MeV. We consider variable final times $\tau$ to explore the quality of the T-QA approach. 
We compile the resulting quantum circuit using Qiskit~\cite{Qiskit2024}, targeting the native gate set \{CNOT, $ R_Z $, $ X $, Hadamard\}.

\section{Results}
\label{Results}
\subsection{Quasiparticle pairing encoding performance}

Before discussing the  results of our quantum annealing simulations, we evaluate the quality of the quasiparticle Hamiltonian $H_Q$ by comparing its ground state properties with those of $H_\text{NSM}$ solved in full configuration interaction (FCI) benchmark simulations. We quantify the performance of $H_Q$ using the relative error of the energy $\Delta_Q E$, and the fidelity, $F_Q$, defined as
\begin{align}
\Delta_Q E&=|E_{Q}-E_\text{FCI}|/|E_\text{FCI}|\,,  \\ 
F_Q&=|\braket{\Psi_Q|\Psi_\text{FCI}}|^2\,,
\end{align}
where $\ket{\Psi_Q}$ and $\ket{\Psi_\text{FCI}}$ are the ground states of $H_Q$ and $H_\text{NSM}$, respectively, with corresponding ground-state energies $E_Q$ and $E_\text{FCI}$. 

Figure~\ref{fig:1} presents our results for these figures of merit for several nuclei across  the $sd$ shell, taking the USDB interaction~\cite{PhysRevC.74.034315} as $H_\text{NSM}$. We observe that oxygen isotopes show excellent agreement between $H_Q$ and FCI, with relative energy errors (left panel) $\Delta_Q E \le 2 \cdot 10^{-2}$ and fidelities (right panel) in the range $F_Q \ge 0.95$. We note, moreover, that for semimagic isotopes with $2$ nucleons or holes of the same species, $H_Q$ reproduces the exact FCI results. This is the case for $^{18}$O --studied in Ref.~\cite{PhysRevC.109.064305}-- and
$^{26}$O in Fig.~\ref{fig:1}, as well 
as $^{18}$Ne and $^{26}$Ar, not shown in the figure because isospin symmetry guarantees the same results for nuclei at both sides of the $N_n=Z_p$ line.

Moving from the oxygen isotopes toward the isospin-symmetric region ($N_n=Z_p$), we find that both the energy and the fidelity accuracy systematically decrease, with $1-F_Q \in [0.15,0.78]$ and $\Delta_Q E \in [2 \cdot 10^{-2},0.22]$. We take this as an indication of the growing contribution from non-quasiparticle configurations in the exact FCI ground state. In fact, the worse results in terms of both energy and fidelity occur for $N_n=Z_n$, especially $^{20}$Ne, $^{24}$Mg and $^{28}$Si, where nuclear ground states are deformed~\cite{Elliot,Frycz:2024clx}, and thus governed by the quadrupole --not pairing-- interaction. The accuracy of our encoding also improves when dealing with more neutron-rich nuclei, as observed by changing $N_n$ while keeping $Z_p$ fixed. Indeed, pairing is responsible for relatively more important quantum correlations in more neutron-rich systems~\cite{Tichai:2022bxr,Perez-Obiol:2023wdz}. 

Figure~\ref{fig:1} also shows a nontrivial relationship between energy accuracy (left panel) and fidelity (right panel). In oxygen isotopes, $\Delta_Q E$ scales roughly with $1 - F_Q$, indicating a direct correlation between energy error and infidelity. In contrast, for silicon isotopes, relatively small energy errors ($\Delta_Q E = 0.06, 0.04, 0.02$) correspond to low fidelities ($1 - F_Q = 0.74, 0.49, 0.34$). This highlights the importance of testing additional ground-state properties besides the energy, as approximate calculations may reach solutions that do not correspond to the real ground state of the system.
We also note that the performance of the quasiparticle pairing encoding is different in particle-hole symmetric nuclei. For example, $^{20}$Ne, with $N_n=Z_p=2$, has an infidelity of $1-F_Q=0.64$, whereas the corresponding particle-hole symmetric isotope, $^{36}$Ar, with $N_n=Z_p=10$, has $1-F_Q=0.42$. This is because the nuclear structure of particle-hole symmetric nuclei depends on the nuclear Hamiltonian. Indeed, $^{20}$Ne is more deformed than $^{36}$Ar and therefore less suited for our quasiparticle mapping.  

%

\begin{figure}[t]
    \includegraphics[width=1\linewidth]{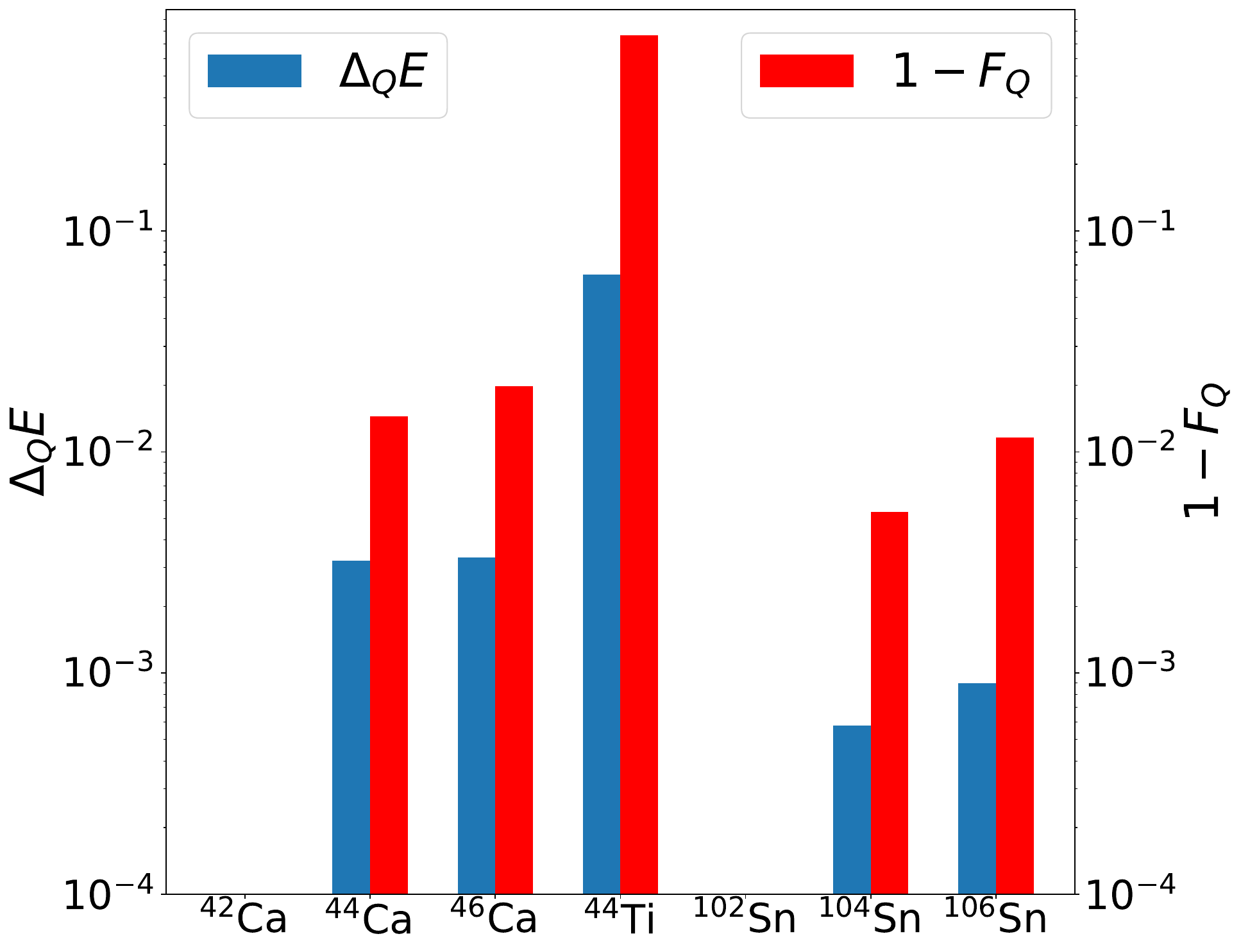}
    \caption{Energy relative error $\Delta_Q E$ (blue bars, left scale) and infidelity $1-F_Q$ (red bars, right scale) compared to FCI results, for Ca isotopes and $^{44}$Ti in the $pf$ shell, and Sn isotopes. The $^{42}$Ca and $^{102}$Sn results are below the scale.}
    \label{fig:2}
\end{figure}

Figure~\ref{fig:2} shows the energy relative error and fidelity of calcium isotopes and $^{44}$Ti in the $pf$ shell, solved for the GXPF1A Hamiltonian~\cite{PhysRevC.69.034335}, as well as tin isotopes in a configuration space including $gdsh$ orbitals using the GCN5082 Hamiltonian~\cite{PhysRevC.80.054311}. The analysis of these results is entirely consistent with those presented for the $sd$ shell. The two-valence-nucleon systems $^{42}$Ca --the focus of Ref.~\cite{PhysRevC.109.064305}-- and $^{102}$Sn are solved exactly, just like $^{18}$O. Besides these nuclei, calcium isotopes show small errors, $\Delta_Q E \le 3 \cdot 10^{-3}$ and high fidelities, $1-F_Q \le 0.02$. For $^{44}$Ti, with $N_n=Z_p=2$, the accuracy of our method drops significantly, with $1-F_Q=0.76$. Just as for the $sd$ shell, moving away from semimagic nuclei degrades the quality of our quasiparticle approach.


For tin isotopes, the fidelities are large, $F_Q \ge 0.99$, and the energy agreement with the FCI benchmark is even better than for oxygen and calcium. These results confirm that the accuracy of the quasiparticle encoding approximation does not diminish in heavier systems. On the contrary, $\Delta_QE$ and $F_Q$ systematically improve for heavier semimagic nuclei. 
Notably, our approach can efficiently encode within the NSM heavy nuclei such as $^{102,104,106}$Sn, and beyond, which 
to our knowledge have not yet been implemented in the literature.
%

\begin{figure}[t]
    \centering
    \includegraphics[width=1\linewidth]{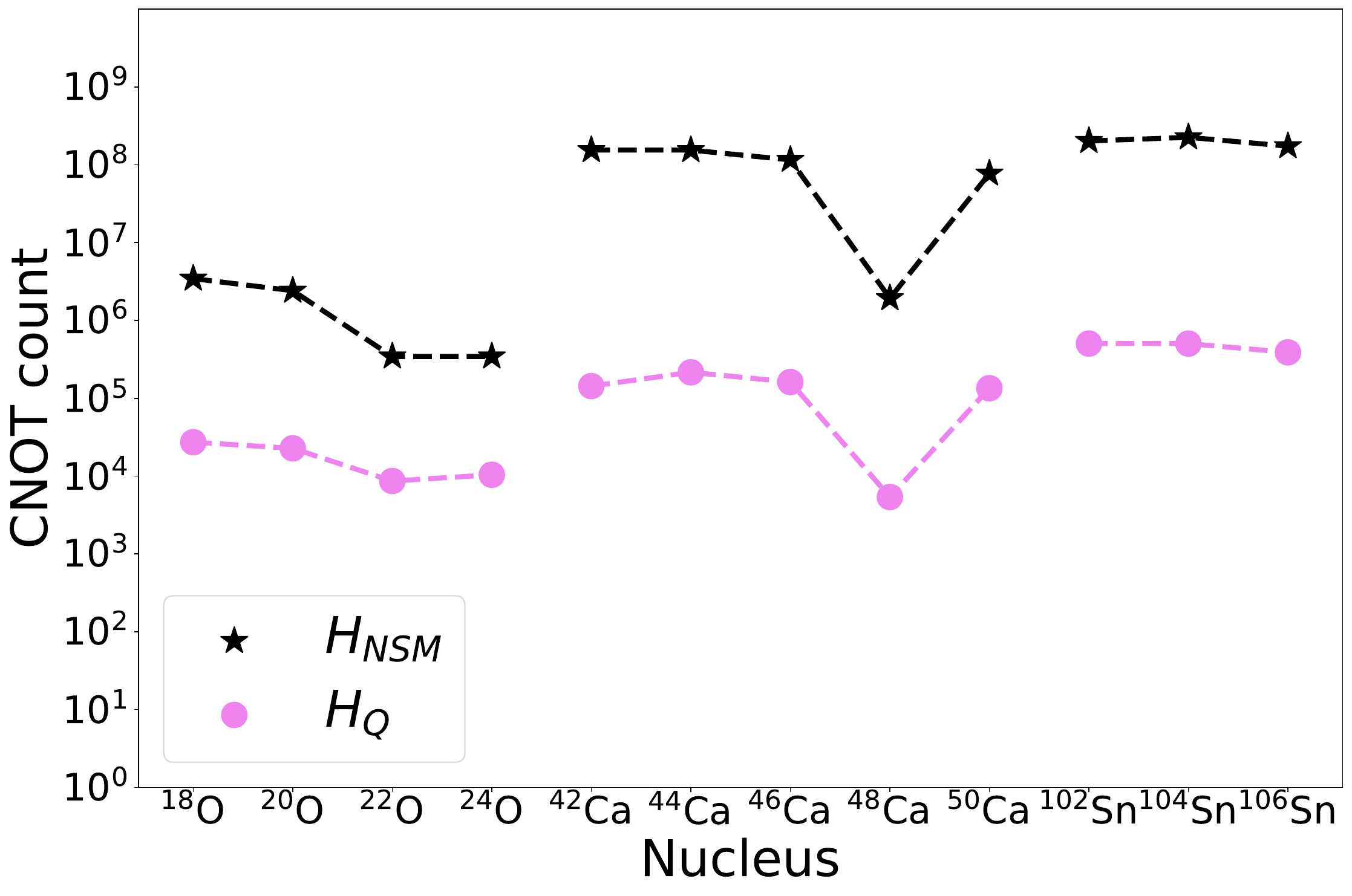}
    \caption{Number of CNOT gates required in the trotterized quantum-annealing circuits for $H_\text{NSM}$ (black stars) and $H_Q$ (violet circles), needed to reach the same fidelity for the ground state of each nucleus (see text for details).}
    \label{fig:4}
\end{figure}

\subsection{Quantum annealing simulations}
We assess the computational advantage of the quasiparticle mapping by simulating adiabatic time evolution within a T-QA framework. We evaluate the CNOT cost per unitary step for 
$U(\Delta t + t, t)$
with $\Delta t$ and the trotterization order chosen to minimize trotterization errors. We then estimate the total number of required steps through classical simulation via matrix multiplication. We compare quantum annealing simulations of $H_\text{NSM}$ (with JW mapping) and $H_Q$ (with the quasiparticle mapping), matching final-state fidelities. For example, for $^{20}$O, the evolution time $\tau$ for $H_\text{NSM}$ is chosen such that $F(\tau)=|\braket{\Psi_\text{FCI}|\Psi (\tau)}|^2=0.95$, the maximum fidelity achievable with $H_Q$ (see the $1-F_Q$ value for $^{20}$O in Fig.~\ref{fig:1}). For the T-QA using $H_Q$, we choose $\tau$ so that the fidelity with respect to the exact $H_Q$ ground state is $F(\tau)=|\braket{\Psi_{Q}|\Psi (\tau)}|^2\ge 0.99$. This guarantees common final-state fidelities for both methods. We follow a similar strategy in the other semimagic nuclei of Fig.~\ref{fig:4}, where the fidelity requirements on the calculations are in all cases higher than for $^{20}$O (see the $1-F_Q$ values Figs.~\ref{fig:1} and~\ref{fig:2}).

Figure~\ref{fig:4} highlights that the quasiparticle approach reduces the number of CNOT gates by $2-3$ orders of magnitude compared to the standard fermion encoding. This suggests a substantial advantage for the practical implementation of the T-QA protocol in a quantum device. While Fig.~\ref{fig:4} focuses on cases with $F_Q \ge 0.95$, the computational benefit of the quasiparticle encoding extends to lower-fidelities as well. 

Remarkably, our results indicate a mild scaling --if any-- of the number of needed CNOT gates as a function of the number of quasiparticle modes in the configuration space. This finding has already been observed for nucleon modes in Ref.~\cite{costa2024quantum}, but over a narrower range of nuclei. Also, Fig.~\ref{fig:4} shows that the trend of the CNOT count as a function of the nucleus is very similar for $H_Q$ and $H_\text{NSM}$. For example, the maximum and the minimum requirement of CNOTs corresponds to $^{22}$O and $^{104}$Sn in both cases. This suggests that the T-QA protocol for $H_Q$ closely resembles the one for $H_\text{NSM}$~\cite{costa2024quantum}, with comparable gaps and structure of the levels. 

\section{Conclusion}
\label{Conclusion}

We investigate a method that, by mapping nucleon modes into quasiparticle pairs with opposite projections, reduces the computational complexity of quantum algorithms to obtain NSM ground states. 
We construct the projected quasiparticle Hamiltonian, $H_Q$, to lowest order and benchmark its performance in reproducing the ground state of the NSM Hamiltonian, $H_\text{NSM}$, for several $sd$- and $pf$-shell nuclei, as well as tin isotopes. Our results show that for semimagic nuclei $H_Q$ provides an excellent approximation, with high fidelity and low relative energy error. Moreover, in these cases we observe a significant computational advantage in the complexity compared to a standard JW encoding of $H_\text{NSM}$. Measured in CNOT gates, our approach reduces the total number up to $2-3$ orders of magnitude. Furthermore, the accuracy of the results and its computational complexity are comparable for lighter oxygen and for heavier tin isotopes. 

However, for general open-shell nuclei whose nuclear structure is not dominated by quasiparticle pairs, the accuracy of our approach is significantly reduced, especially because it does not reach high fidelities. In order to address this limitation, we plan to investigate two possible avenues. First, one could explore hybrid fermion–quasiparticle representations \cite{del2025hybrid}, where selected nucleon modes remain in the fermionic basis. This would compromise the simplicity of our novel approach while capturing additional nuclear correlations relevant for open-shell nuclei. Second, one may add perturbative corrections to $H_Q$ using the Brillouin–Wigner formalism \cite{Hubac_2000,brillouin3,brillouin1,HubacWilson2009_BWMethods,brillouin2} to account for non-quasiparticle contributions via effective interactions. 

Overall, our approach offers a promising pathway toward efficient quantum simulations of medium-to-heavy nuclei, balancing accuracy and resource savings while paving the way for future extensions to general, more complex nuclear systems.

\vspace{0.3cm}
\emph{Note added:} while preparing this manuscript, the authors became aware of the preprint in Ref.~\cite{arxiv2509.20642}, presenting a similar hard-core boson mapping for the NSM in the context of VQE simulations.

\section*{Acknowledgements}

This work is financially supported by 
MCIN/AEI/10.13039/501100011033 from the following grants: PID2021-127890NB-I00, PID2023-147112NB-C22, CNS2022-135529 and CNS2022-135716 funded by the
“European Union NextGenerationEU/PRTR”, and 
and CEX2024-001451-M  to the “Unit of Excellence Mar\'ia de Maeztu 2025-2031” award to the Institute of Cosmos Sciences; and by the Generalitat de Catalunya, through grant 2021SGR01095. 
This work is also financially supported by the Ministry of Economic Affairs and Digital Transformation of the Spanish Government through the QUANTUM ENIA project call – Quantum Spain project, and by the European Union through the Recovery, Transformation and
Resilience Plan – NextGenerationEU within the framework of the Digital Spain 2026 Agenda.

\bibliographystyle{elsarticle-num}
\bibliography{biblio.bib} 

\end{document}